\documentclass[aps,prappl,reprint,a4paper,superscriptaddress,floatfix,amsmath,amssymb,amsfonts,noshowpacs,longbibliography]{revtex4-2}
\usepackage{newtxtext,newtxmath}
\usepackage[T1]{fontenc}
\usepackage[utf8]{inputenx}
\usepackage{graphicx}
\usepackage[dvipsnames]{xcolor}
\usepackage{soul}
\usepackage[textwidth=17.5cm,textheight=23.5cm,verbose,pdftex]{geometry}
\usepackage[pdftex]{hyperref}

\hypersetup{pdfauthor={Georg Hoffmann}, bookmarksnumbered=true, pdftitle={BSO growth}, colorlinks, citecolor=blue, linkcolor=blue, urlcolor=blue}

\usepackage{floatrow}
\usepackage{csquotes}

\DeclareMathAlphabet{\mathcal}{OMS}{cmsy}{m}{n}

\begin{document}
\title{Adsorption-controlled plasma-assisted molecular beam epitaxy of LaInO$_3$ on DyScO$_3$(110): Growth window, strain relaxation, and domain pattern }
\author{Georg Hoffmann}
\email[Electronic mail: ]{Hoffmann@pdi-berlin.de}
\affiliation{Paul-Drude-Institut für Festkörperelektronik, Leibniz-Institut im Forschungsverbund Berlin e.\,V., Hausvogteiplatz 5--7, 10117 Berlin, Germany}
\author{Martina Zupancic}
\author{Detlef Klimm}
\author{Robert Schewski}
\author{Martin Albrecht}
\affiliation{Leibniz-Institut für Kristallzüchtung, Max-Born-Straße 2, 12489 Berlin, Germany}
\author{Manfred Ramsteiner}
\author{Oliver Bierwagen}
\email[Electronic mail: ]{Bierwagen@pdi-berlin.de}
\affiliation{Paul-Drude-Institut für Festkörperelektronik, Leibniz-Institut im Forschungsverbund Berlin e.\,V., Hausvogteiplatz 5--7, 10117 Berlin, Germany}
\begin{abstract}
  We report the growth of epitaxial LaInO$_3$ on DyScO$_3$(110) substrates by adsorption-controlled plasma-assisted molecular beam epitaxy (PA-MBE). The adsorption-controlled growth was monitored using line-of-sight quadrupole mass spectrometry. In a thermodynamics of MBE (TOMBE) diagram, the experimental growth window was found to be significantly narrower than the predicted one. We found the critical thickness for strain relaxation of the LaInO$_3$ layer (lattice mismatch $\approx-4\%$) to be of 1~nm using in-situ RHEED analysis. Substrate and film possess an orthorhombic crystal structure which can be approximated by a pseudo-cubic lattice.  X-ray-diffraction (XRD) analysis revealed the pseudo-cube-on-pseudo-cube epitaxial relationship of the LaInO$_3$ films to the DyScO$_3$ substrates. This relation was confirmed by transmission electron microscopy (TEM), which further resolved the presence of rotational orthorhombic domains - the majority of which have coinciding $c$-axis with that of the substrate. Raman spectroscopy further confirmed the presence of a LaInO$_3$ layer. Our findings open up the possibility for 2-dimensional electron gases at the MBE-grown heterointerface with BaSnO$_3$. 
\end{abstract}

\maketitle

\section{Introduction}
\label{sec:introduction}

The family of complex oxides comprises dielectric, semiconducting, superconducting, ferromagnetic, or ferroelectric materials. Their common, perovskite, crystal structure provides the basis for combining these oxides and their respective properties epitaxially to form (multi)-functional heterostructures. 
Among these multi functional heterostructures, non-polar SrTiO$_3$ (STO) interfaced with polar perovskites such as LaAlO$_3$\cite{janotti2012,mannhart2008,ohtomo2004}, LaTiO$_3$\cite{ohtsuka2010} or GaTiO$_3$\cite{moetakef2011,mikheev2015} can be considered as the working horse for the evaluation of a 2-dimensional electron gas (2-DEG) realized by a polar discontinuity at the polar/non-polar interface.\\
However, room-temperature mobilities of these 2-DEGs as well as doped STO films are below 10~cm$^2$/Vs due to the high electron phonon interaction,\cite{mikheev2015} impeding any possible transistor application.\\

In search for other suitable channel materials, having the chance for future applications, non-polar 
La-doped BaSnO$_3$ (BSO) single crystals, with reported room-temperature electron mobilities up to 320 cm$^2$/Vs ,\cite{kim2012a} surpassed the mobility values of STO by almost two orders of magnitude.
The mobilities achieved in BSO single crystals are also the highest values within the perovskites family up to today, making BSO an excellent channel material for a 2-DEG.\cite{krishnaswamy_2016}

For the realization of such a 2-DEG, polar LaInO$_3$ (LIO) has been predicted to have a conduction band offset of 2.06~eV towards non-polar BSO that can confine a maximum possible 2-DEG electron density of almost 2$\times 10^{14}$ per cm$^2$ due to the polar discontinuity at the LIO/BSO interface.\cite{krishnaswamy_2016}\\
So far LIO single crystals have been grown from the melt,\cite{galazka2021} and LIO thin films on lightly La-doped BSO thin films by pulsed laser deposition (PLD). These LIO/BSO:La heterostructures resulted in the formation of a 2-DEG at their interface,\cite{pfutzenreuter2022} but also rotational domains in the LIO film due to the epitaxial mismatch between orthorhombic LIO and cubic BSO.\cite{zupancic2020}
It is also assumed that the crystal quality is reduced because of the high particle energies in PLD, that cause a higher concentration of point defects such as cation vacancies, interstitials and anti-site defects compared to low energy deposition methods such as molecular beam epitaxy (MBE). \cite{nunn2021,schmid2009} 
While different flavors of molecular beam epitaxy (MBE) have demonstrated the adsorption-controlled growth of BSO thin films with the highest quality and electron mobility\cite{paik_2017,raghavan2016,prakash_2015}, no growth  of LIO by MBE has been reported to date. \\
Here, we demonstrate the adsorption-controlled growth of LIO thin films by plasma-assisted MBE on DyScO$_3$ (DSO) (110) substrates. The growth window, lattice-mismatch-related layer relaxation, and rotational domain pattern are discussed. 


\section{Basic considerations} 
\label{sec:basics}
\begin{figure}[t]
\centering
\includegraphics[width=1\textwidth]{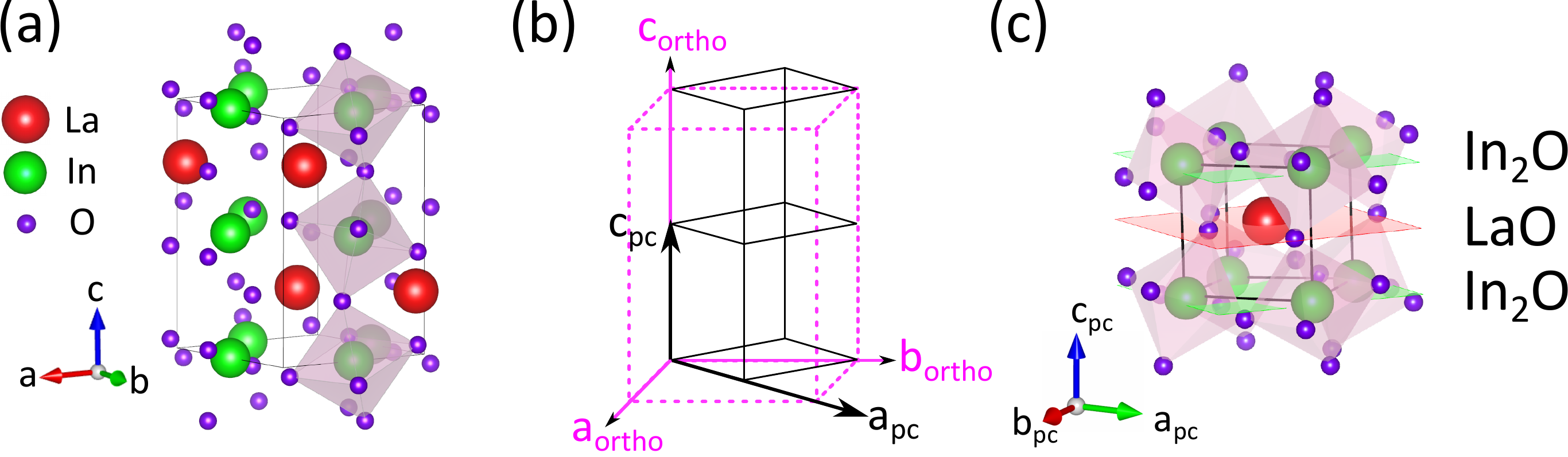}
\caption{(a) Orthorhombic LIO unit cell with three octahedra formed by the oxygen atoms highlighted in purple. (b) Sketch of the relationship of the orthorhombic (pink) and related $1\times1\times2$ pc (black) LIO unit cell.(c) LIO pc unit cell. The oxygen atoms are positioned at the corners of the purple octahedra. The In$_2$O and LaO planes are indicated as purple and green planes, respectively. }\label{Fig_LIO_unit_cell}
\end{figure}
\begin{table}[b]
\caption{Lattice mismatch in $\%$ of DSO and LIO with respect to their aligned pc directions (given as row and column heading).}
\begin{tabular}{|c|c|c|}
\hline 
 & LIO $c_{\text{pc}}$ & LIO $a$/$b_{\text{pc}}$ \\ 
\hline 
DSO $c_{\text{pc}}$ & -3.93 & -4.33 \\ 
\hline 
DSO $b_{\text{pc}}$ & -4.04 & -4.46 \\ 
\hline 
\end{tabular}
\label{Tab_lattice}
\end{table}

\subsection{Crystal structure}
As shown in Fig.~\ref{Fig_LIO_unit_cell}(a), LIO possesses an orthorhombic unit cell with lattice parameters $a=0.57227$~nm, $b=0.5938$~nm, and $c=0.82143$~nm.\cite{galazka2021} The In atoms are surrounded by oxygen atoms in a tilted octahedral arrangement. LIO forms a coinciding (100)-oriented pseudo-cubic (pc) unit cell in its orthorhombic (110)- and (001)-orientation as schematically shown in Fig.~\ref{Fig_LIO_unit_cell}(b). The lattice constants $a_{\text{pc}}= b_{\text{pc}}=\frac{1}{2}\sqrt{a^2+b^2}$~=~0.41234~nm, and $c_{\text{pc}}=c/2$~=~0.41072~nm represent the pc orientations (100)$_{\text{pc}}$, (010)$_{\text{pc}}$, and (001)$_{\text{pc}}$, respectively. In addition, the angle between $a_{\text{pc}}$ and $b_{\text{pc}}$ of 87.6$^{\circ}$ deviates from the 90$^{\circ}$ of an exact cube.\cite{zupancic2020} Note, that for multiple unit cells, $a_{\text{pc}}$ and $b_{\text{pc}}$ can be assumed to be identical and are therefore summarized into $a_{\text{pc}}$.
Depending on the orientation of the c$_{\text{pc}}$ direction, three different rotational domains contribute to the same cubic $ \lbrace$100$\rbrace_{\text{pc}}$ orientation. A detailed discussion of the octahedral arrangement of the LIO with respect to a single unit cell and its domain pattern on cubic BSO(100) is given elsewhere.\cite{zupancic2020}
As shown in Fig.~\ref{Fig_LIO_unit_cell}(c), the pc cell of LIO still results in an alternating stacking of polar LaO~(+1) and InO$_2$~(-1) planes due to the charge transfer of 0.5 electrons per unit cell from the LaO to the InO$_2$ below and above.\cite{krishnaswamy_2016}

DyScO$_3$ (DSO) also possesses an orthorhombic crystal structure with lattice constants $a$~=~0.5442417~nm, $b$~=~0.5719357~nm, and $c$~=~0.7904326~nm,\cite{schmidbauer2012} which allows for a definition of a pc unit cell with $a_{\text{pc}}= b_{\text{pc}}=\frac{1}{2}\sqrt{a^2+b^2}$~=~0.39475~nm, and $c_{\text{pc}}=c/2$~=~0.39522~nm. Since we grow our LIO films on DSO(110) substrates, the DSO surface is spanned by $b_{\text{pc}}$ and $c_{\text{pc}}$.
There are three possible orientations of LIO to grow pc-on-pc on the DSO(110) (one with $c_{\text{pc}}$ in out-of-plane and two variants with $a_{\text{pc}}$ in out-of-plane direction, but in-plane rotated with respect to each other by 90$\thinspace ^{\circ}$
The calculated lattice mismatches of the LIO film to the DSO substrate with respect to the possible orientations of the pc cells are shown in Tab.~\ref{Tab_lattice}, and suggest compressive strain in all cases.

For the following discussion and analysis of the samples, the pc notation is indicated by the subscript "pc" whereas no subscripts are used in the orthorhombic notation.
\subsection{Adsorption-controlled growth and TOMBE diagram}
\begin{figure}[t]
\centering
\includegraphics[width=0.9\textwidth]{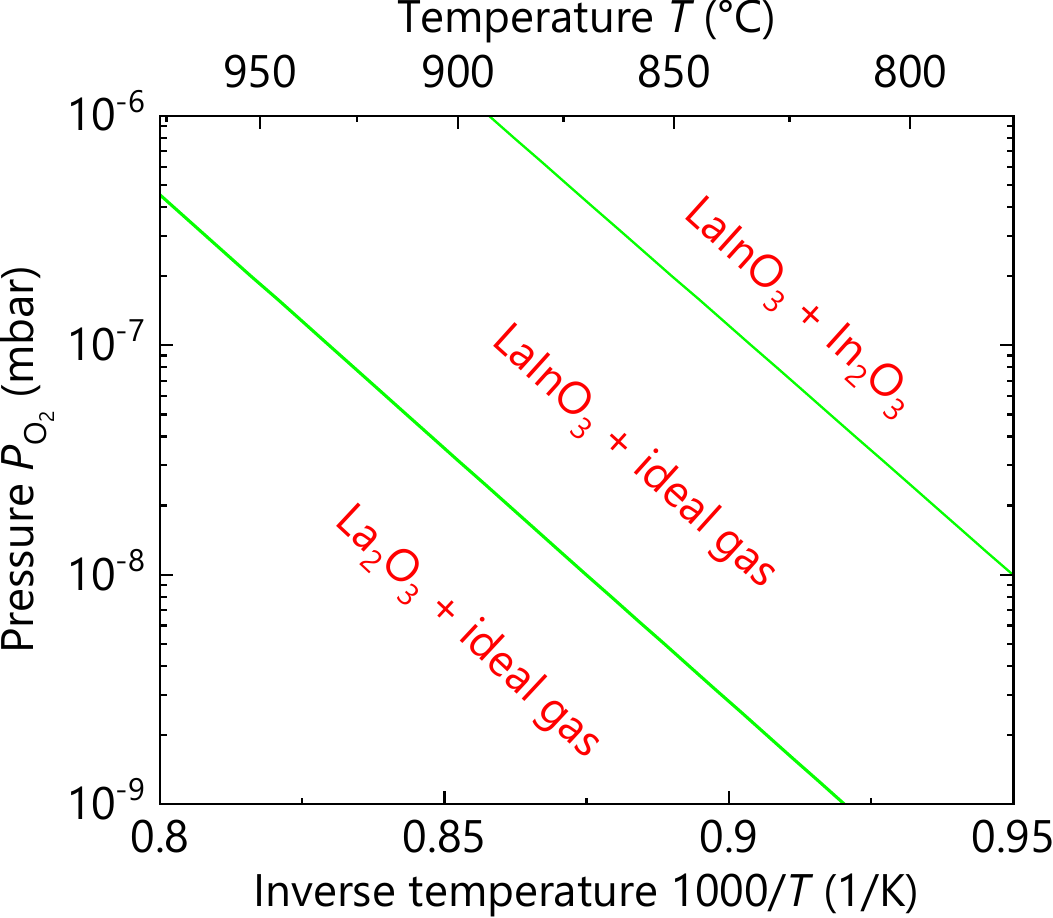}
\caption{Thermodynamics of MBE (TOMBE) diagram for LIO growth showing the theoretically expected solid equilibrium phases, LIO~+~In$_2$O$_3$, LIO, and La$_2$O$_3$ (green lines). The ideal gas mainly consists of the volatile In$_2$O.}\label{Fig_Ellingham}
\end{figure}

Possessing two different cations (A and B), a common challenge for the MBE-growth of complex oxides (ABO$_3$) is the realization of the proper A:B-cation stoichiometry of 1:1.
As long as one cation can desorb from the film surface at a given growth temperature, the film stoichiometry can self-adjust as the adsorption of this cation is controlled by the flux of the other cation, the O-flux, and the thermodynamically stable film composition. 
Suboxides provide an additional desorption channel that enables adsorption-controlled growth even for cations with prohibitively-low vapor pressure, since suboxides often have a significantly higher vapor pressure than their parent cations. The related suboxides can be formed either at the source \cite{hoffmann2020,hoffmann2021} or intermediately at the growth front.\cite{Vogt2018} 
This "adsorption-controlled" approach has been used, for example, to grow BaSnO$_3$. \cite{paik_2017, prakash_2017a}

\begin{figure}[b]
\centering
\includegraphics[width=0.9\textwidth]{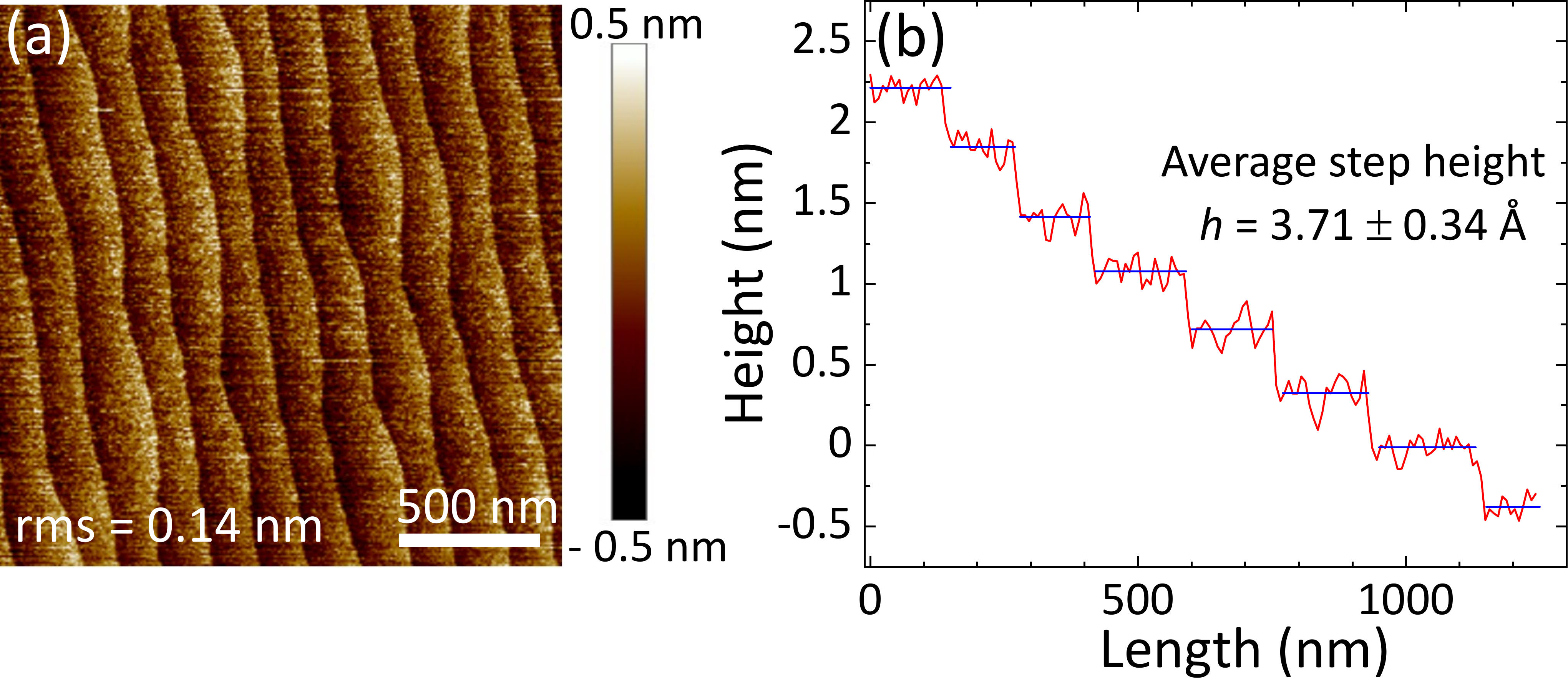}
\caption{(a) AFM image of a DSO 110 oriented (100 oriented in pc notation) substrate after annealing treatment. (b) Height profile of the step terrace structure of (a). The average step height $h$ is within the theoretically expected lattice distance of 3.95~$\mathring{A}$ with respect to the pc unit cell.}\label{Fig_DSO_sub_01}
\end{figure}
In the case of LIO, the low vapor pressures of La (and its suboxide LaO) result in a sticking coefficient of unity at the growth temperatures accessible to us, whereas In (and its suboxide In$_2$O) possess a high-enough vapor pressure to grant desorption from the growth surface.\cite{hoffmann2021} To delineate the growth conditions that provide an LIO film free of any secondary solid La$_2$O$_3$ or In$_2$O$_3$ phases, a TOMBE diagram shown in Fig.~\ref{Fig_Ellingham} was computed using the FactSage\texttrademark \space thermochemical software package.\cite{factsage} 
Such a diagram has been introduced by Nair $et$ $al.$ in the context of adsorption-controlled MBE of ruthenates,\cite{nair2018} and shows the formed solid phases as function of temperature and oxygen partial pressure. 
For the calculation of the LIO phase, the heat capacity $c_p = a + bT + d/T^2$, with $a = 136.8228$, $b = 5.8324 \times 10^{-3}$, and $d = -1.847436 \times 10^6$ [in J/(mol K)], and formation enthalpy $\Delta H = -13.9$~kJ/mol of bulk LIO were determined experimentally.\cite{klimm2021}
The values were added to the FactSage\texttrademark -database. The TOMBE diagram was calculated assuming an In excess and for an oxygen pressure of 10$^{-6}$~mbar to match our MBE conditions in terms of In-flux and O$_2$ background pressure. The three different regions in the diagram can be explained as follows: (I) at low temperatures or high oxygen pressures In$_2$O$_3$ is formed next to LaInO$_3$ due to the chosen In excess. (II) at higher temperatures or lower oxygen pressures no In$_2$O$_3$ is formed and the In$_2$O suboxide desorbs from the surface leaving solid LaInO$_3$ behind due to its lower formation enthalpy. (III) at very high temperatures, also the LaInO$_3$ is decomposed and La$_2$O$_3$ is formed.

\section{Experimental methods} \label{sec:experiment}

\begin{figure}[t]
\centering
\includegraphics[width=0.9\textwidth]{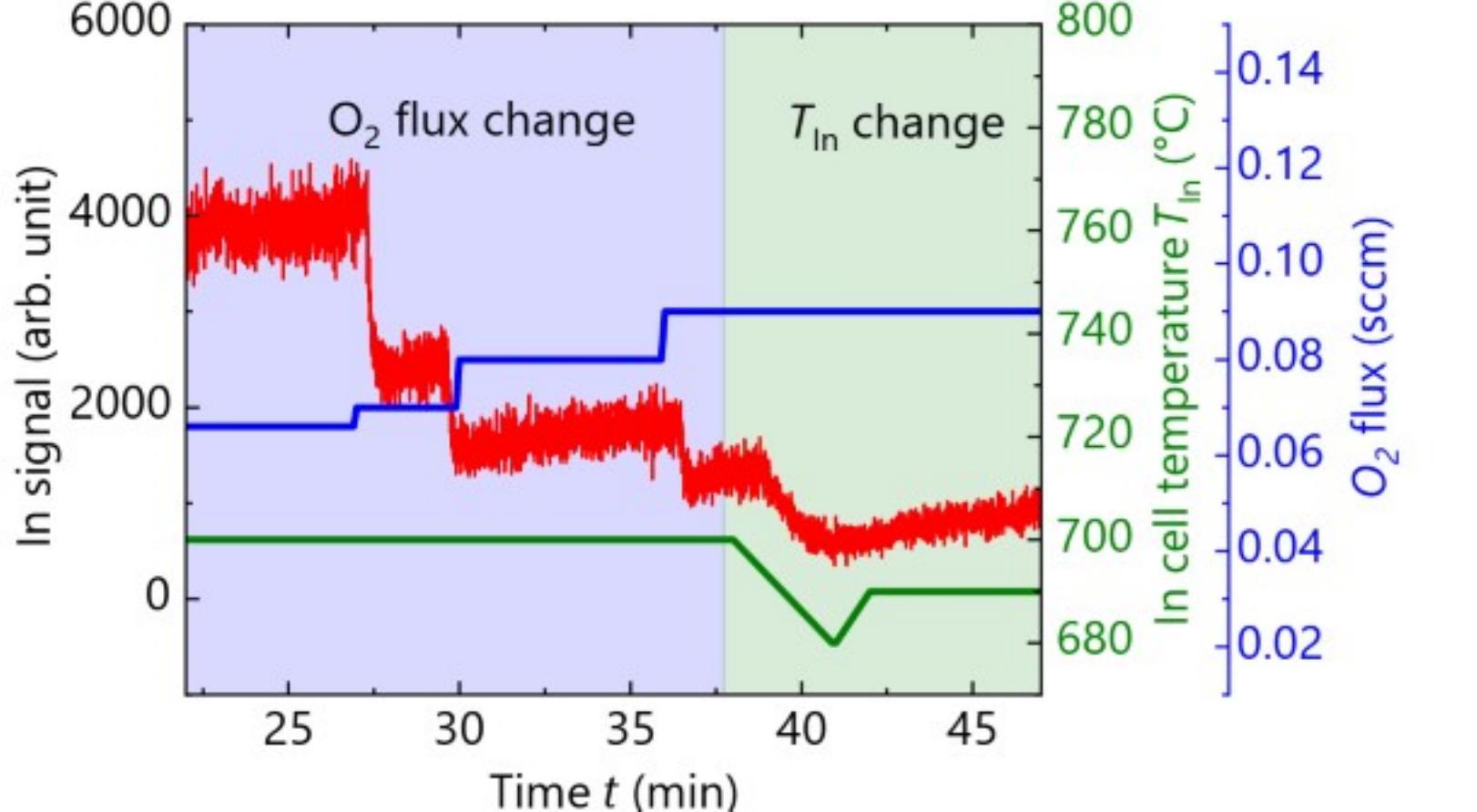}
\caption{Measured In- or In$_2$O-desorption from LIO growth front for varying O$_2$ flux (blue shaded area) and varying In cell temperature (green shaded area).}\label{Fig_QMS}
\end{figure}

We grew the LIO films on DSO(110) substrates from CrysTec GmbH [(100)$_{\text{pc}}$ oriented]. The DSO substrates were annealed in a quartz tube furnace at 1050$^{\circ}$C for 6 h in flowing oxygen gas at atmospheric pressure [O$_2$ flux at $\approx$ 250 standard cubic centimeters per minute (sccm)] in order to achieve a well defined surface. Fig.~\ref{Fig_DSO_sub_01}(a) shows an AFM image of the annealed DSO surface reflecting the unintentional offcut of the wafer. The height profile of the step terrace structure in Fig.~\ref{Fig_DSO_sub_01}(b) reveals monolayer steps, indicating single site terminated surface for the subsequent LIO growth. We assume that this site is ScO$_2$ (B-site) according to the chosen annealing parameters.\cite{dirsyte2010}
All samples were grown using an rf-plasma source operated at 200~W and used O$_2$ flow rates ranging from 0.07~-~0.14~sccm which corresponds to an oxygen partial pressure of 2.2$\times 10^{-7}~-~4.7\times10^{-6}$~Torr in the growth chamber. In order to keep the plasma ignited at low oxygen fluxes, 0.2 sccm Ar were added to the gas flow using a second mass-flow controller, a method previously demonstrated by Tolstova $et$~$al.$ \cite{tolstova2015}
The La and In cell temperatures of $T_{\text{cell}}^{\text{La}}$~=~1550$^{\circ} \thinspace$C and $T_{\text{cell}}^{\text{In}}$~=~700$^{\circ} \thinspace$C
result in beam equivalent pressure (BEP) values of 3$\times 10^{-8}$~and $\approx 3.3 \times 10^{-8}$~mbar, respectively, measured by a nude ion gauge at the substrate position without oxygen flux. Under these conditions, the derived La flux of $\approx 2.8 \times 10^{13}$cm$^{-2}s^{-1}$ determines the growth rate of 1.1~nm/min, while In is provided in excess with a flux of $5.7 \times 10^{13}$cm$^{-2}s^{-1}$.
The substrate temperature (925$^{\circ}$C if not indicated otherwise) was measured by a pyrometer.

The surface of the samples during growth was analysed $in-situ$ by reflection high energy electron diffraction (RHEED) at an electron energy of 20 keV. The RHEED images were recorded by CCD camera with a resolution of 1024x1024 pixel. Further, a quadrupole mass spectrometer (QMS), that is mounted line-of-sight to the substrate, monitors species that desorb from the sample surface.

Fig.~\ref{Fig_QMS} shows the desorbing In$_2$O-flux (measured as Indium fragment) by line-of-sight QMS. In the blue shaded region, the desorbing In flux decreases with increasing O-flux, as expected from the TOMBE diagram, indicating adsorption control by the O-flux. In the green shaded area, the In signal follows the In cell temperature since less In is being desorbed when less In is provided. \\

The samples were investigated $ex-situ$ by atomic force microscopy (AFM) using a Dimension Edge AFM from Bruker in the peak-force tapping mode, and by X-ray diffraction (XRD) with a X’pert pro MRD from Philips PANalytical using a 1~mm detector slit. High-resolution Scanning transmission electron microscopy (STEM) was performed with FEI Titan 80-300
operating at 300 kV and equipped with a Fischione high-angle annular dark-field detector (HAADF) and a highly brilliant cathode (X-FEG). The semi-convergence angle was tuned to 9 mrad and the semi-acceptance angle of the detector was set to 35 mrad. Cross-sectional TEM samples were prepared along the ($\overline{1}$10) [(010)$_{\text{pc}}$] lattice direction of the DSO
substrate by tripod polishing and argon ion-milling at liquid nitrogen temperature. Ar+ ion-milling was done
by a precision ion polishing system (PIPS) at beam energies from 4.0 to 0.2 keV

The Raman spectroscopic measurements were performed in the backscattering configuration at room temperature using the intracavity frequency-doubled output of an Ar+ ion laser at 244~nm (5.08 eV) for optical excitation. The incident laser light was focused by a microscope objective onto the sample surface. The backscattered light was collected by the same objective, spectrally dispersed by an 80-cm spectrograph (LabRam HR Evolution, Horiba/Jobin Yvon) and detected by a liquid-nitrogen-cooled charge-coupled device (CCD) without analysis of the polarization. 


\section{Results and Discussion}\label{sec:results}

 \subsection{Strain relaxation in LIO films}\label{subsec_strain_relax}
 
 \begin{figure*}[t!]
\centering
\includegraphics[width=0.9\textwidth]{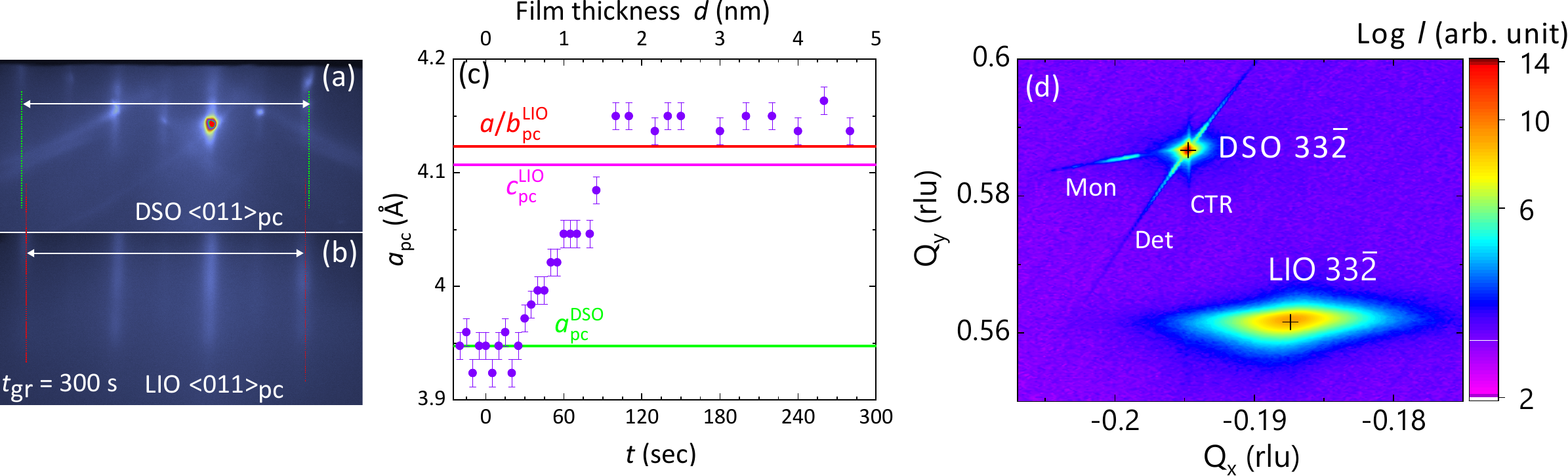}
\caption{(a) RHEED pattern of a DSO(100)$_{\text{pc}}$ substrate along the <011>$_{\text{pc}}$ azimuth. (b) RHEED pattern of a LIO(011)$_{\text{pc}}$ film (intended growth direction) along the <110>$_{\text{pc}}$ azimuth after 300 sec growth time (i.e. 5 nm thickness). The green and red dashed lines in (a) and (b) are guides to the eye to demonstrate the change in the measured lattice constant (white arrows). 
(c) In-plane lattice parameter of the LIO film by $in-situ$ RHEED as a function of the growth time. The colored lines are the respective pc lattice constants of DSO and LIO. (d) Reciprocal space map of the 33$\bar{2}$ reflection of the grown sample at 0.09 sccm (see discussion of LIO growth window). The
crystal truncation rod, the monochromator streak, and the detector streak are labeled by CTR, Mon, and
Det, respectively. The black crosses show the theoretically expected reflection position indicating a fully relaxed LIO layer. Note that the sample used for the analysis (a)~-~(c) was grown under similar conditions as the one shown in (d).}\label{Fig_RHEED_lattice}
\end{figure*}

The growth of LIO on DSO was $in-situ$ monitored by RHEED. Fig.~\ref{Fig_RHEED_lattice}(a) and~\ref{Fig_RHEED_lattice}(b) show the <011>$_{\text{pc}}$ azimuth of the DSO(100)$_{\text{pc}}$ substrate and the nominal LIO(100)$_{\text{pc}}$ layer after 300 seconds, respectively. The LIO in-plane lattice parameter was calculated from the analysis of the RHEED patterns [i.e. measuring the distance between the streaks as shown in Fig.~\ref{Fig_RHEED_lattice}(a) and~\ref{Fig_RHEED_lattice}(b), which is inversely proportional to the in-plane lattice parameter] in order to follow the strain relaxation of the LIO films on DSO. In Fig.~\ref{Fig_RHEED_lattice}(c) we see an increase of the lattice parameter after 30 seconds ($\approx$~1~monolayer) of film growth from the DSO lattice constant (solid green line) towards the expected LIO lattice constant (solid red line). After 2~nm of LIO growth the determined lattice constant indicates a fully relaxed LIO film. Such a relaxation after 1~nm of film growth or less is also known from BSO films that were grown on SrTiO$_3$ substrates.\cite{prakash_2015}\\
A reciprocal space map (RSM) of the 33$\bar{2}$ reflection of a 200~nm thick LIO film on a DSO substrate is shown in Fig.~\ref{Fig_RHEED_lattice}(d). The black crosses mark the theoretically expected position of the DSO and LIO reflections indicating a fully relaxed LIO layer despite the broadened LIO peak in agreement with RHEED analysis. 
\subsection{LIO growth window}
\label{subsec_growth}

\begin{figure*}[t]
\centering
\includegraphics[width=1.0\textwidth]{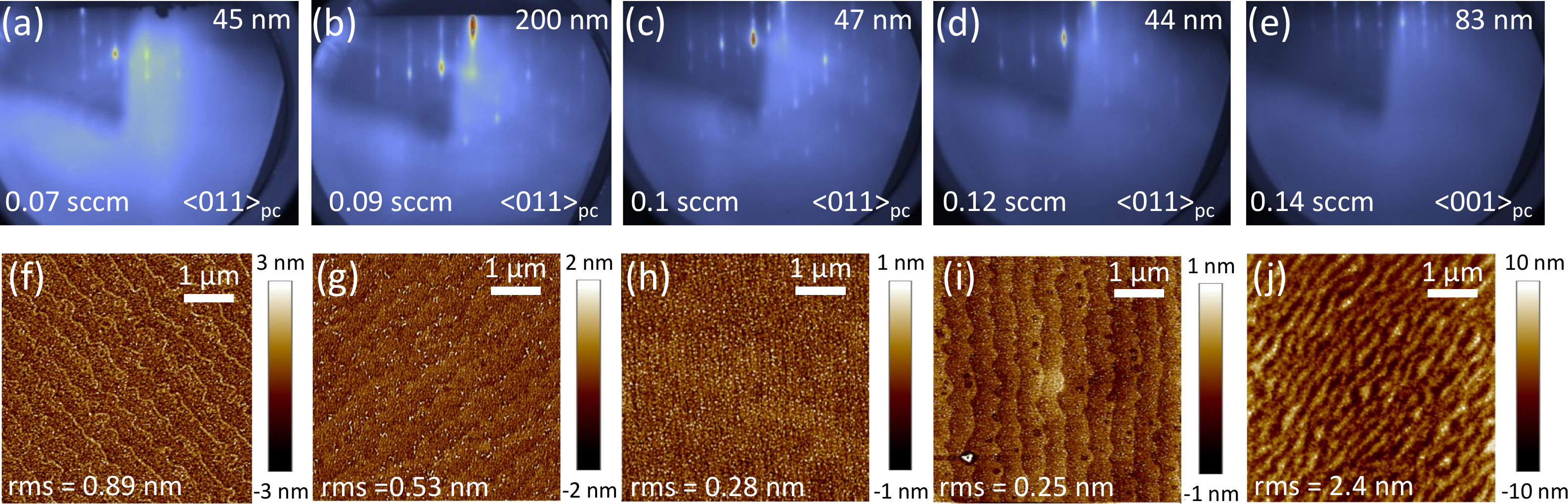}
\caption{RHEED images of the <011>$_{\text{pc}}$ azimuth (a)-(d) and <001>$_{\text{pc}}$ azimuth (e) for different oxygen fluxes ranging from 0.07 to 0.14~sccm. All azimuths are labelled according to the pc notation. (f)-(j) corresponding AFM images of the LIO surface. }\label{Fig_RHEED_AFM}
\end{figure*}

For the determination of the growth window, we evaluated the RHEED patterns, AFM images, and XRD diffractograms as a function of oxygen flux.

Fig.~\ref{Fig_RHEED_AFM}(a)-(e) show RHEED patterns of the LIO(100)$_{\text{pc}}$ surface along the <011>$_{\text{pc}}$ (a)-(d) and <001>$_{\text{pc}}$ (e) azimuth obtained at the final film thickness as indicated. For oxygen fluxes between 0.09 and 0.12 sccm, a streaky RHEED pattern with 1'st and 2'nd Laue circles can be observed whereas for 0.07 and 0.14 sccm a more pronounced intensity modulation of the streaks in the RHEED patterns indicates the transition to three dimensional growth. Note that Fig.~\ref{Fig_RHEED_AFM}(e) shows the RHEED pattern along the <001> azimuth since along this direction the effect of the intensity modulation of the streaks is more pronounced than in <011> direction. Fig.~\ref{Fig_RHEED_AFM}(f)-(j) show the corresponding AFM images of the LIO surface. For the samples grown at 0.09, 0.1 and 0.12 sccm the surface shows the step terrace structure of the underlying substrate even for a 200~nm thick LIO film [Fig.~\ref{Fig_RHEED_AFM}(g)] and root mean square (rms) values in the range of 0.25~-~0.53~nm. Also for the films grown at 0.07 and 0.14 sccm the terrace structure of the substrate still seems to have an influence on the growth of the LIO film. The rms values of 0.89 and 2.4~nm, however, are significantly higher than for the other samples.

\begin{figure}[t]
\centering
\includegraphics[width=1.0\textwidth]{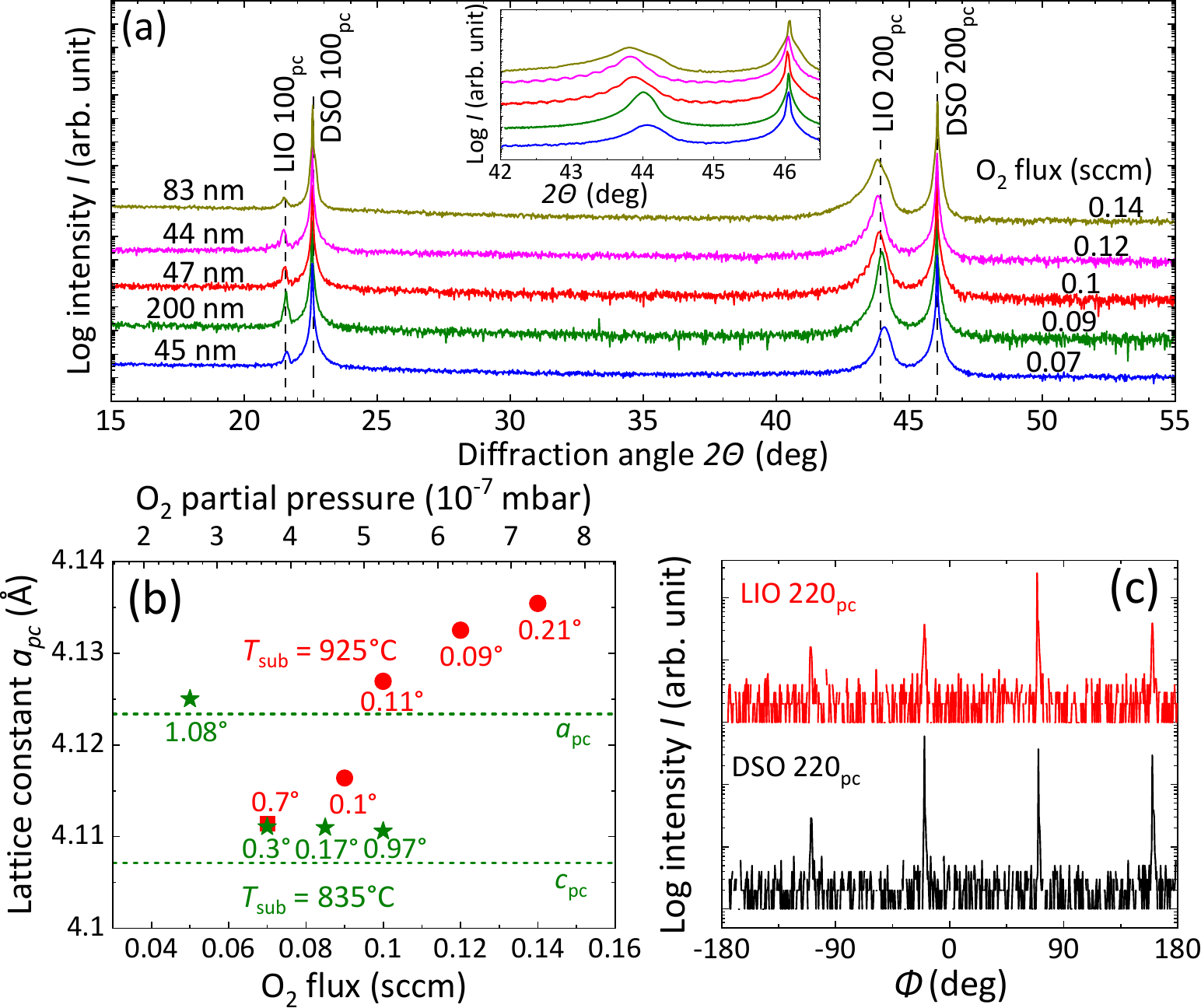}
\caption{(a) Wide-angle $\omega$-2$\Theta$ scan of the LIO films on DSO. Inset: Enlarged area of the LIO 200$_{\text{pc}}$ reflection. (b) red discs: lattice constants calculated from LIO 200$_{\text{pc}}$ reflection in the inset of (a). Other icons: LIO grown at different La and In fluxes, and $T_{\text{sub}}$~=~835$\thinspace ^{\circ}$C. The corresponding full width at half maximum (FWHM) values derived from LIO 200$_{\text{pc}}$ rocking curves are written next to the icons. The green dashed lines mark the expected lattice constant assuming single domain LIO growth along the $c_{\text{pc}}$ and $a_{\text{pc}}$. (c) Out of plane $\Phi$ scan of the sample grown at 0.1~sccm indicating epitaxial relationship between the substrate and the film.
}\label{Fig_X-ray}
\end{figure}

In Fig.~\ref{Fig_X-ray}(a) the symmetric 2$\Theta$~-~$\omega$ XRD scan (using a 1~mm detector slit) along the out-of-plane direction reveals the successful growth of LIO for all applied fluxes. No additional phases (e.g., In$_2$O$_3$ or La$_2$O$_3$) were observed within the detection limit of XRD. Solely for a LIO film grown at significantly lower substrate temperature of 730$\thinspace ^{\circ}$C and $P_{\text{O}_2}$~=~7.8$\times 10^{-8}$~mbar, an additional phase was observed (not shown). We identified this phase as In$_2$O$_3$ by its 222 reflection in agreement with the predicted In-rich regime in Fig.~\ref{Fig_Ellingham}. \\
The thickness of the LIO films was determined by the Laue fringes of the LIO 200$_{\text{pc}}$ reflection shown in the inset for thin films and extrapolated to thicker films assuming the same growth rate for all films. Further, the inset of Fig.~\ref{Fig_X-ray}(a) reveals a shift of the 200$_{\text{pc}}$ peak position towards lower angles with increasing oxygen flux. Consequently, the derived out-of-plane lattice constants of the LIO films with respect to the oxygen flux in the pc notation shown in Fig.~\ref{Fig_X-ray}(b) (indicated by red squares) also increase with increasing oxygen flux. However, when growth conditions such as substrate temperature are changed (as it was done for the samples labelled by green asterisks), no shift of the lattice constant is observed. \\
As it will also be shown in the section \ref{subsec_domains}, the presence of rotational domains allow for an upper and lower boundary of expected LIO lattice constants assuming total $a_{\text{pc}}$ orientation (upper boundary) or total $c_\text{pc}$ orientation (lower boundary) of the LIO films as indicated by the green dashed lines in Fig.~\ref{Fig_X-ray}(b).
However, due to the use of the 1~mm slit on the detector size, the out-of-plane lattice constants of the samples shown in Fig.~\ref{Fig_X-ray}(b) can only be determined to a value of 4.12~\AA~ with an uncertainty of~$\approx$~0.01~\AA. For higher precision of the LIO lattice constant, and for a statement on the preferred domain formation as a function of oxygen, high-resolution XRD would be necessary.\\


\begin{figure}[t]
\centering
\includegraphics[width=1.0\textwidth]{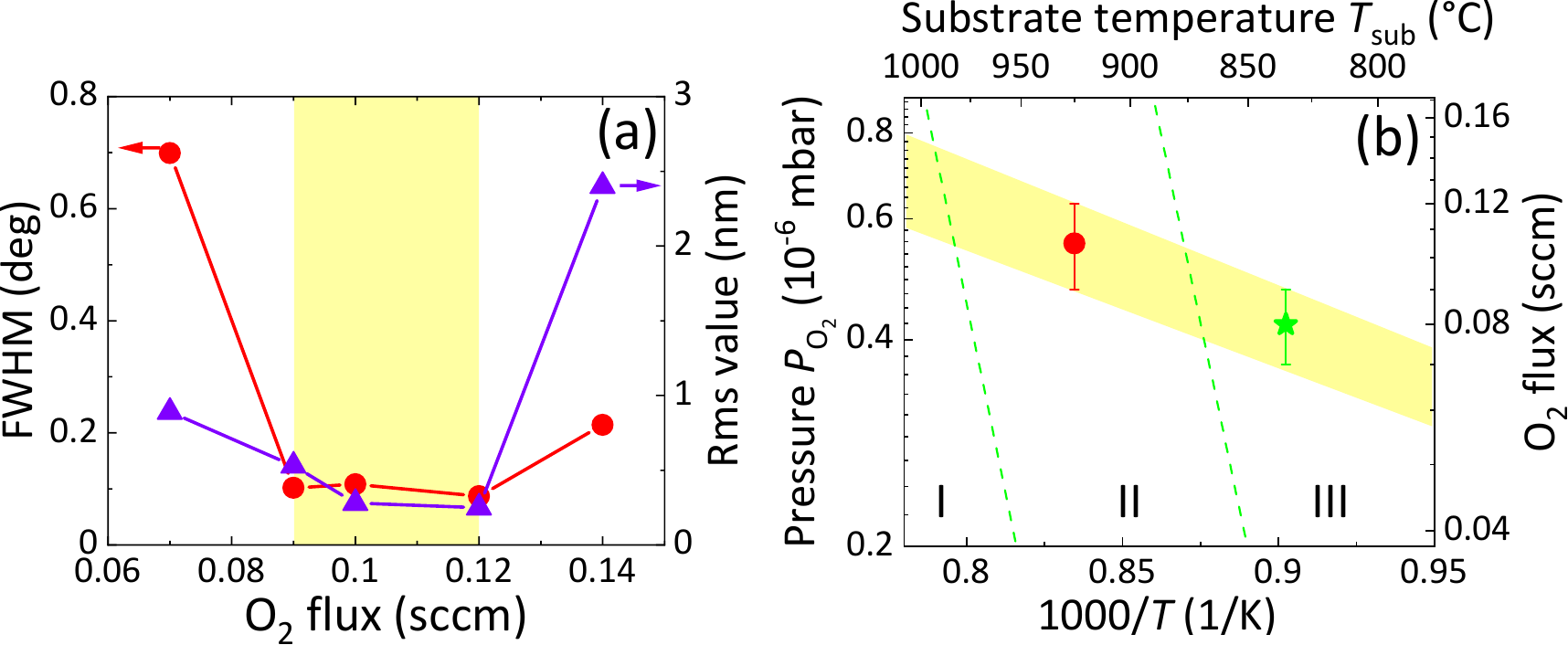}
\caption{(a) FWHM of $\omega$-rocking curve values from XRD analysis, (left axis) and rms roughness values (right axis) from AFM analysis of the grown samples shown in Fig.~\ref{Fig_RHEED_AFM}(f)-(j). The yellow shaded area marks the empirical growth window with respect to the applied fluxes. (b) TOMBE diagram of
experimentally derived growth window. The two data points indicate the growth conditions at which high quality LIO films were grown with respect to FWHM and rms values as shown in (a). The yellow region indicates the extrapolated growth
window according to the derived data. The green dashed lines indicate the calculated TOMBE diagram from Fig.~\ref{Fig_Ellingham} with La$_2$O$_3$ containing (I), pure LIO (II), and In$_2$O$_3$ containing (III) phases.}\label{fig_exp_vs_theory}
\end{figure}
An empirical growth window at a fixed $T_\text{sub}$~=~925$\thinspace ^\circ$~C was determined by identifying the range of oxygen flux that minizes both, the FWHM of the LIO 200$_{\text{pc}}$  XRD $\omega$-rocking  curve and the rms surface roughness of the LIO layer. Both these quantities are shown in Fig.~\ref{fig_exp_vs_theory}(a) as red discs (XRD) and purple triangles [roughness determined from the AFM images shown in Fig.~\ref{Fig_RHEED_AFM}(f)~-~(j)]) and the yellow shaded area marks the growth window. The LIO 200$_{\text{pc}}$ FWHM values in the range of 0.09$^{\circ}$~-~0.15$^{\circ}$ are higher by a factor of 2~-~3 compared to the ones achieved for BSO \cite{paik_2017}, likely related to the presence of the rotational domains of the pc lattice (see discussion in section \ref{subsec_domains}).\\
In Fig.~\ref{fig_exp_vs_theory}(b) the experimentally derived growth windows at two different $T_{\text{sub}}$ of 925$\thinspace ^{\circ}$C and 835$\thinspace ^{\circ}$C are indicated by the error bars of the data points. Consequently, the LIO growth window can be extrapolated through the data points as indicated by the yellow shaded area. \\
The derived LIO growth window as function of substrate temperature and oxygen flux is strikingly different from the predicted TOMBE diagram indicated by the green dashed lines. \\

\subsection{Epitaxial relation, domain structure, and dislocations of the LIO films}\label{subsec_domains}


\begin{figure}[t]
\centering
\includegraphics[width=0.9\textwidth]{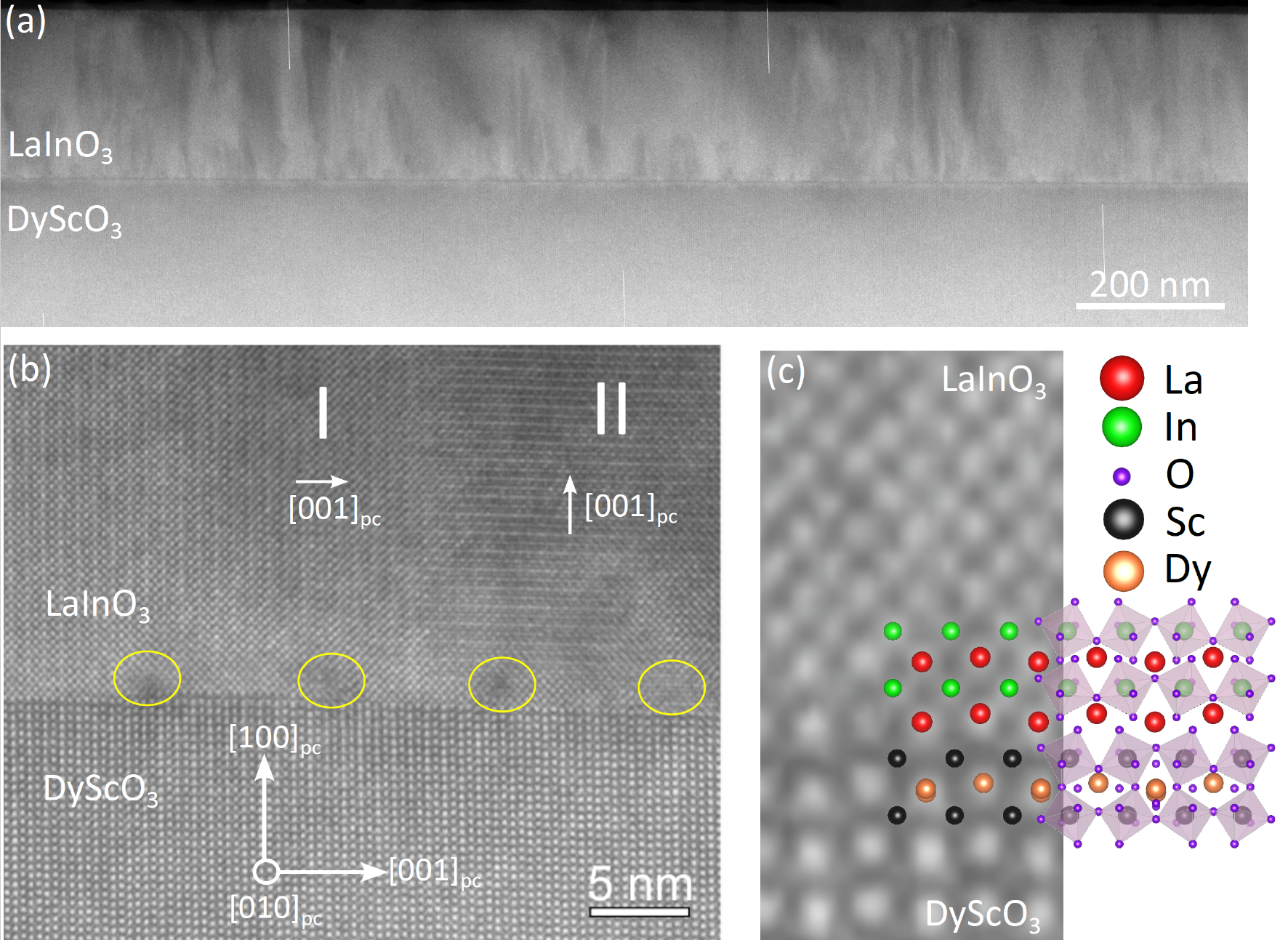}
\caption{(a) STEM HAADF image of the DSO/LIO heterostructure. (b) High resolution STEM HAADF image of the DSO/LIO interface. Rotational domains (I and II) are visible within the LIO film. The yellow circles indicate dislocations according to the lattice mismatch between film and substrate. (c) Magnified area of DSO/LIO interface overlapped with corresponding atomic models of DSO and LIO build in VESTA,\cite{Momma2011} and according to the growth conditions described in \ref{sec:experiment}.} \label{Fig_TEM_01}
\end{figure}

The $\Phi$-scan along the DSO and LIO 220$_{\text{pc}}$  reflection shown in Fig.\ref{Fig_X-ray}(c) in addition to the coinciding 200$_{\text{pc}}$ reflections of LIO film and DSO substrate [Fig.\ref{Fig_X-ray}(a)] reveals the pc-on-pc epitaxial relation between DSO and LIO. 

To get a better understanding on the rotational domains of the the orthorhombic structure, that we cannot resolve by XRD, TEM investigations on the sample grown at 0.09 sccm and 925$\thinspace ^{\circ}$C were performed. Fig.~\ref{Fig_TEM_01}(a) shows a STEM HAADF overview of the layer. The layer is characterised by a high density of dislocations, visible as brightness variations under the imaging conditions used here. They start at the interface and  penetrate the layer.
At higher magnification, the misfit dislocations can be directly seen at the interface [marked by yellow circles in the Fig.~\ref{Fig_TEM_01}(b)]. The measured spacing of $\approx$~8.9~nm between two misfit dislocations corresponds well with the expected dislocation density for the mismatch between DSO and LIO in the given orientation, and for full relaxation (in agreement with our RHEED observations and RSM).
A closer look at Fig.~\ref{Fig_TEM_01}(b) shows the layer to consist of domains, revealed by the specific contrast pattern. The contrast patterns labelled I and II correspond to LIO domains with $a_{\text{pc}}$ and $c_{\text{pc}}$, respectively, parallel to the surface normal of the (100)$_{\text{pc}}$ oriented DSO substrate as described in work by Zupancic \textit{et} al. \cite{zupancic2020} for LIO grown on BSO. The formation of differently oriented domains have also been observed in orthorhombic SrRuO$_3$ and CaRuO$_3$ films on cubic substrates.\cite{proffit2008a,jiang1998a}


\begin{figure}[t]
\centering
\includegraphics[width=0.9\textwidth]{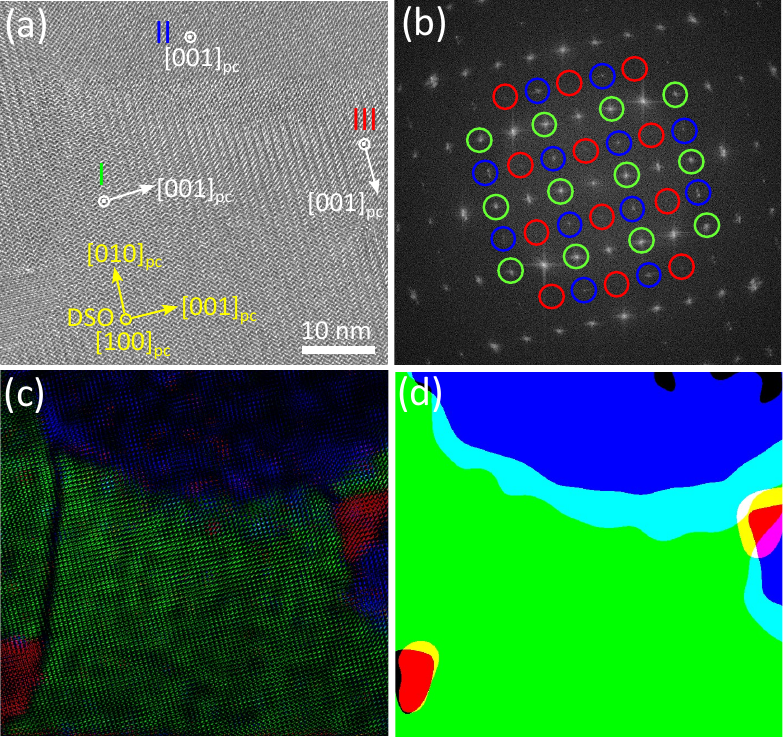}
\caption{(a) HRTEM image of plan-view sample showing the LIO surface. The yellow arrows indicate the DSO substrate orientation. The white arrow indicates the $c_\text{pc}$-direction of the domain I (marked by green), II (oriented out-of-plane, blue), and III (red). (b) Fast Fourier transformation of HRTEM image where the reflexes from differently oriented domains are marked by circles using the same color code as in (a). (c) RGB composite image of Bragg filtered domains. (d) RGB threshold image of three different type of domains.}\label{fig_domain_analysis}
\end{figure}

For a statistical analysis of the different LIO domains, we analyze the plan-view sample. Fig.~\ref{fig_domain_analysis}(a) shows a part of the analyzed LIO surface along the (100)$_\text{pc}$ surface normal of the DSO substrate. Different domain orientations, labelled I, II, and III are marked with green, blue and red, respectively. Domain II (blue) corresponds to $c_\text{pc}$ parallel to the surface normal of the DSO substrate, whereas domain I (green) and III (red) indicate $a_\text{pc}$ out-of-plane oriented LIO with $c_\text{pc}$ parallel and perpendicular, respectively, to $c_\text{pc}$ of the DSO substrate.
To make different orientations distinguishable, we perform Bragg filtering as described in Zupacic $et$ $al.$ \cite{zupancic2020} using fast Fourier transformation [Fig.~\ref{fig_domain_analysis}(b)]. In this way, each filtered image shows only domains with the same orientation [Fig.~\ref{fig_domain_analysis}(c)]. After binarization of filtered images, we extract the percentage of each orientation. To better visualize this, we show in Fig.~\ref{fig_domain_analysis}(d) an RGB image combined of the processed binary images. In total, a sample area of 0.12 $\mu$m$^2$ was analyzed. Our analysis shows that $a_\text{pc}$ oriented domains with coinciding $c_\text{pc}$ directions of substrate and film (I, green) cover 78.0~±~9.5$\%$ of the surface, while those rotated 90$^\circ$ (III, red) cover only 0.2~±~0.4$\%$. The $c_\text{pc}$ oriented domains (II, blue) occupy 21.8~±~9.5~$\%$ of LIO film. 

Our findings indicate a strong preference for the alignment of the c-axis of orthorhombic LIO film to that of the orthorhombic DSO(110) substrate (domain I, green color) whereas the LIO domain with c-axis rotated in plane by 90$^{\circ}$ (domain III, red color) is almost completely suppressed. The derived domain distribution differs distinctly from that of orthorhombic LIO grown on cubic BSO, where the three domains are more evenly distributed.\cite{zupancic2020} Even though our results are not in agreement with lowest strain values listed in Tab.~\ref{Tab_lattice}, an impact of the orthorhombic substrate on the domain distribution is assumed.

\subsection{Raman spectroscopy}

\begin{figure}[t]
\centering
\includegraphics[width=1.0\textwidth]{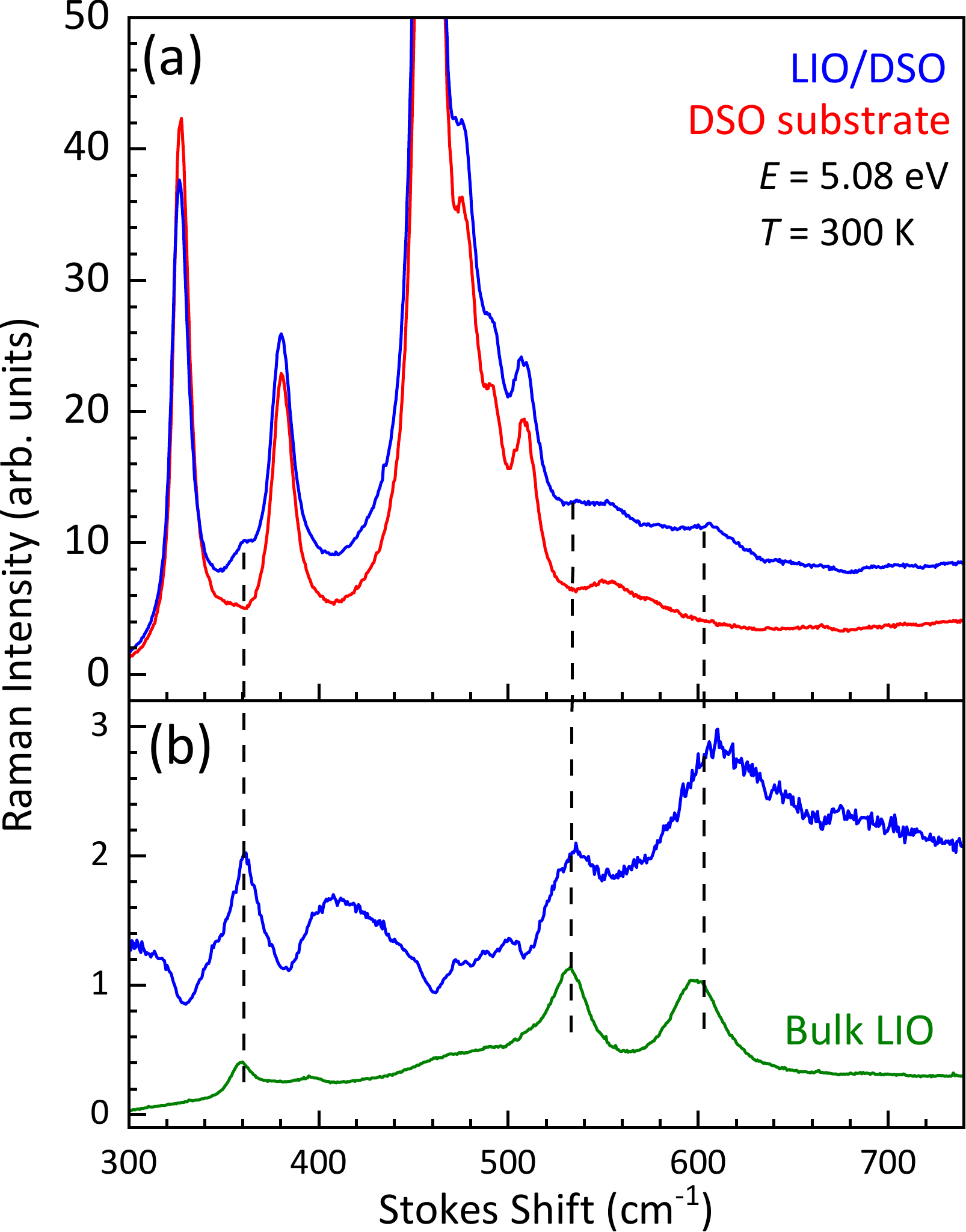}
\caption{(a) Raman spectra of LIO film grown under optimized conditions on a DSO 100$_{\text{pc}}$ substrate (blue line) together with that of a bare DSO substrate (red line) excited at a photon energy of 5.08~eV. (b) Ratio spectrum obtained by dividing the LIO/DSO spectrum by that of the DSO substrate (blue line). The spectrum of a melt-grown bulk LIO 100$_{\text{pc}}$ sample is shown for comparison (green line). The black dashed lines indicate the positions of the major Raman peak observed for bulk LIO under optical excitation at 5.08 eV. }\label{fig_raman}
\end{figure}

The Raman spectrum of a 100~nm thick LIO film grown on a DSO(100)$_{\text{pc}}$ substrate is shown in Fig.~\ref{fig_raman}(a) together with the spectrum of a bare DSO substrate. Despite the optical excitation (at 5.08~eV) clearly above the optical bandgaps of LIO (4.35 and 4.39~eV, see Ref.~\cite{galazka2021}), the optical probing depth in the LIO/DSO sample is large enough to result in a dominating Raman signal from the DSO substrate. In order to extract more clearly the contributions of LIO from the LIO/DSO spectrum, we divided the latter one by the one from the DSO substrate.\cite{ju2017} The resulting ratio spectrum is shown in Fig.~\ref{fig_raman}(b) together with the spectrum of a bulk LIO 100$_{\text{pc}}$ sample recorded under the same conditions. We have detected the same major Raman peaks around 360, 535, and 600 cm$^{-1}$ for bulk LIO with different surface orientations, however, with somewhat different relative intensities. In fact, the ratio spectrum exhibits three Raman peaks which coincide with those of bulk LIO and have about the same peak widths. This finding confirms the growth of orthorhombic LIO films on DSO substrates under the chosen growth conditions with a crystal quality comparable to that of melt-grown bulk crystals. Note that the identification of the LIO Raman modes for the grown LIO film is rather challenging due to the used DSO substrate. Therefore, for future Raman studies, the use of other substrates is recommended.


\section{Summary and conclusion}\label{sec:summary}
In summary, we have grown LIO thin films on DSO(110) substrates by PA-MBE. Measurements of the desorbing flux as function of growth parameters confirms an adsorption-controlled growth regime, that allows desorption of excess In. A TOMBE diagram for the growth has been theoretically derived. The experimentally determined growth window for high-quality LIO as function oxygen flux for two different substrate temperatures was found to only qualitatively agree with that of the TOMBE diagram. XRD, TEM, and Raman spectroscopy confirmed the formation of an epitaxial LIO film. RHEED, XRD, and TEM show, that the lattice mismatch on the order of -4$\%$ resulted in a film relaxation within the first 2~nm by the formation of misfit dislocations at the LIO/DSO interface, as well as threading dislocations that penetrate the entire LIO film. While both substrate and film are orthorhombic, their epitaxial relation is shown to be pc-on-pc with 90$^\circ$ rotational domains of the film. In contrast to LIO films grown on BSO with quite evenly distributed rotational domains,\cite{zupancic2020}  the majority of the LIO domains in the present work have a coinciding orthorhombic $c$-axis with that of the DSO substrate (both oriented in-plane) while also domains with $c$-axis oriented out-of-plane are present. It is assumed that 001 oriented substrates with a smaller lattice mismatch may lead to a reduction of rotational domains \cite{suyolcu2021} as well as shutter controlled layer-by-layer growth that in general has been demonstrated to lead to higher crystal quality.\cite{suyolcu2020,haeni2000,eckstein1995,putzky2020a,wrobel2017} 
The present work paves the way for the formation of polarization-discontinuity-doped 2-DEGs with high room temperature electron mobility in LIO/BSO heterostructures grown by MBE. 

\begin{acknowledgments}
We thank Zbigniew Galazka for providing the LIO substrates, Hans-Peter Schönherr and Steffen Behnke for technical support, and Dr. Philipp John and Dr. Jutta Schwarzkopf for critically reading the manuscript. This work was performed in the framework of GraFOx, a Leibniz-ScienceCampus partially funded by the Leibniz association. G.H. and M.Z. gratefully acknowledge financial support from the Leibniz association under Grant No. K74/2017.\newline
\end{acknowledgments}

\begin{center}
\textbf{Data Availability}
\end{center}

The data that support the findings of this study are available from the corresponding author upon reasonable request.


%

\end{document}